\documentclass[twocolumn,preprintnumbers,amssymb,amsmath,superscriptaddress,letterpaper,nofootinbib]{revtex4}[12pt]
\usepackage{times,graphics}
\usepackage[colorlinks]{hyperref}

\usepackage{graphicx}
\usepackage{dcolumn}
\usepackage{bm}
\usepackage{natbib}
\usepackage{pstricks}
\usepackage{epsfig}
\usepackage{epstopdf}
\usepackage{multirow}
\usepackage{amsmath}
\usepackage{lipsum}

\newcommand{\be}{\begin{equation}}
\newcommand{\ee}{\end{equation}}
\newcommand{\bea}{\begin{eqnarray}}
\newcommand{\eea}{\end{eqnarray}}

\makeatletter
\newcommand*{\shifttext}[2]{%
	\settowidth{\@tempdima}{#2}%
	\makebox[\@tempdima]{\hspace*{#1}#2}%
}
\makeatother

\begin{document}

\title{Phenomenological Gravitational Phase Transition: Early and Late Modifications}

\author{Nima Khosravi}
\email{n-khosravi@sbu.ac.ir}
\affiliation{Department of Physics, Shahid Beheshti University, 1983969411,  Tehran, Iran}

\author{Marzieh Farhang}
\email{m\_farhang@sbu.ac.ir}
\affiliation{Department of Physics, Shahid Beheshti University, 1983969411,  Tehran, Iran}

\date{\today}

\begin{abstract}
	In this work we generalize the idea of a gravitational phase transition or GPT at late times \citep{Farhang:2020sij} to allow for a modified gravity scenario  in the early universe as well. 
	The original GPT was shown to  simultaneously relax the $H_0$ and  $\sigma_8$ tensions and {\it Planck} internal inconsistencies.
	However, the primary results from GPT predictions implied disagreement with the data from the baryonic acoustic oscillations (BAO). 
	Here we investigate whether the generalized GPT scenario could address the Hubble tension in the presence of the BAO data as well.  We find that, despite the vastly enlarged gravitational parameter space,  the BAO data strongly prefer the $\Lambda$CDM paradigm with a low inferred Hubble constant, $68.03\pm  0.76  $ km/s/Mpc, in tension with the local $H_0$ measurements from \cite{R19}, $H_0=74.03\pm 1.42$ km/s/Mpc. 
	This failure of the generalized GPT scenario to host the P18-BAO-R19 trio is of significance and noticeably shrinks the space  of
	possible gravitational solutions to the Hubble tension. 
	
\end{abstract}

\maketitle

\section{Introduction and Motivations}
The recent cosmological tensions between early and late time observations in the context of the standard model of cosmology, the so-called $\Lambda$CDM,  have been extensively addressed by many cosmologists in the last few years.
However, a truly convincing  theoretical model for their explanation is still missing. 
The main tension is the well-known $H_0$-tension which is the $\sim\,4.2\sigma$ discrepancy between the local measurements of the Hubble constant \citep{R16,R18,R19} and its inferred value  from the cosmic microwave background (CMB) data  in the 
$\Lambda$CDM framework \citep{pl18}\footnote{For some controversy on the significance of the Hubble tension see \cite{Efstathiou:2021ocp,Camarena:2021jlr}.}. There have also been many attempts for  $H_0$ estimation from other independent observations, with higher $H_0$ values \citep{fre19,hua19,won19,khe20,pes20}.  
Beside the Hubble tension, there is the milder $\sigma_8$ tension with its local measurements impling less clumpiness in matter distribution compared to the CMB-based inferences \citep{pl18}. 
Certain internal inconsistencies are also reported in  {\it Planck} results, specifically in the CMB lensing amplitude and the low- and high-$\ell$ discrepancy. 
If not due to systematics, and if considered statistically significant, theoretical models beyond $\Lambda$CDM are required to explain them. 

Many different models are proposed to address these cosmological tensions \citep[see][for a comprehensive review]{DiValentino:2021izs}. 
The models could in general be categorized into late and early, based on the era of the modification.
The motivation behind the late time modifications is clear and there are many attempts \citep[see, e.g.,][]{DiValentino:2017iww,Zhao:2017cud,Khosravi:2017hfi,yan18,Yang:2018uae,Banihashemi:2018has,Li:2019yem,kee19,Raveri:2019mxg,Yan:2019gbw,dival20,Braglia:2020iik,Gomez-Valent:2020mqn,luc20,DiValentino:2019ffd,DiValentino:2019jae,Banihashemi:2018oxo,Banihashemi:2020wtb}. The early modifications try to reduce the sound horizon at the last scattering surface which results in a higher value for $H_0$ \citep{Poulin:2018cxd,Knox:2019rjx,kee20,Vagnozzi:2021gjh}. There are many other different ideas to approach the $H_0$ tension, such as dark matter \citep{vat19}, primordial magnetic fields \citep{Jedamzik:2020krr} and etc. Neither of these models are consistent with all the datasets, in particular when data from the baryonic acoustic oscillations (BAO) are included \cite{Jedamzik:2020zmd}.

In a recent work \citep{Farhang:2020sij} the authors proposed a phenomenological phase transition in the gravity sector by assuming that the early standard $\Lambda$CDM transits to a modified gravity model both at the background and perturbation levels. We verified that the model could resolve the CMB-lensing, low- and high-$\ell$'s, $\sigma_8$ and $H_0$ tensions simultaneously (at different levels). 
The BAO datasets were not included in the main analysis and parameter estimation of \cite{Farhang:2020sij}. However, the predictions of GPT for BAO observables were shown to disagree with some BAO data points. 
In this work we generalize the model to allow for transition between two general {\it early} and {\it late} gravity models at some intermediate redshift. We will study this model in the presence of BAO datasets. 
These standard rulers provide us with independent observations at low redshifts, where a transition in the gravity sector was shown to relax the tensions between CMB and local Hubble measurements. 
On one hand, BAO data are expected to break the parameter degeneracies and improve parameter measurements. 
On the other hand, the enhanced space of gravitational degrees of freedom at early times may have the potential to accommodate both  BAO data and local $H_0$ values. 
This is the  driving motivation of this paper: to explore the huge subset of gravity models effectively modelled by a transition at some redshift, with possible simultaneous deviations from GR  in early and late times, and search for a possible sweet spot where CMB, BAO and local $H_0$ measurements can all co-exist in peace.

\section{The GPT model}\label{sec:model}
The gravitational phase transition, or GPT model, introduced in \cite{Farhang:2020sij}, suggests a transition in the gravitational theory, phenomenologically modelled by a \texttt{"tanh"} function. It was assumed that the early phase of gravitation is governed by the standard Einstein-Hilbert model,  referred to as GR, which forms the gravity model in the $\Lambda$CDM scenario. 
In this work we relax this assumption and study the consequences.

We assume our Universe is isotropic, homogeneous and  flat, where the background evolution is given by  
\begin{eqnarray}
	H^2(z)=H_0^2 \bigg[\Omega_r (1+z)^4+\Omega_m (1+z)^3+\Omega_\Lambda(z)\bigg]
\end{eqnarray}
and  linear scalar perturbations are described by 
\begin{eqnarray}
	ds^2=a^2(\tau)[&-&(1+2\Psi(\tau,\vec{x}))d\tau^2 \nonumber\\
	&+&(1-2\Phi(\tau,\vec{x}))d\vec{x}^2 ]
\end{eqnarray}
in the Newtonian gauge. Here $\Psi$ and $\Phi$ characterize the scalar perturbations and $a(\tau)$ is the scale factor.
In Fourier space the linear Einstein equations can be modified phenomenologically as 
\begin{eqnarray}\label{eq:mu}
	&&k^2 \Psi = -\mu(z)\,4\pi G a^2\,\left[\rho\Delta+3(\rho+P)\sigma\right]\\
	&&k^2\left[\Phi-\gamma(z)\Psi\right]=\mu(z)\,12\pi G a^2\,(\rho+P)\sigma, \label{eq:gamma}
\end{eqnarray}
where $\mu(z)$ and $\gamma(z)$ are responsible for any general redshift-dependent modifications to GR. 
Equations (\ref{eq:mu}) and (\ref{eq:gamma}) can also be written as  
\begin{equation}\label{eq:lensing}
	k^2(\Phi+\Psi)=8\pi G a^2 \,\Sigma(z)\, \rho\, \Delta
\end{equation}
which is particularly useful for studying the gravitational lensing.
In the GPT scenario, we assume a transition in the gravity sector and model the transition through \texttt{"tanh"} function. At the background level we therefore have
\begin{eqnarray}\label{eq:delv}
	\Omega_\Lambda(z)=\Omega^{\rm early}_\Lambda+\Delta_\Lambda\,\frac{1+\tanh\big[\alpha\,(z_{\rm t}-z)\big]}{2}.
\end{eqnarray}
Here $z_{\rm t}$ is the transition redshift and $\alpha$ is the inverse of the transition's width. 
Imposing the flatness condition gives $\Omega_{\rm r}+\Omega_{\rm m}+\Omega_\Lambda=1$, where 
$\Omega_\Lambda\equiv\Omega_\Lambda(z=0)=\Omega^{\rm early}+\Delta_\Lambda[1+\tanh(\alpha z_{\rm t})]/2$. 
Similarly, the modifications to the perturbation equations are characterized by transitions in $\mu(z)$ and $\gamma(z)$,
\begin{eqnarray}\label{eq:pert}
	\mu(z)&=&\mu^{\rm early}+\Delta_{\mu} \frac{1+\tanh\big[\alpha\,(z_{\rm t}-z)\big]}{2}, \nonumber \\
	\gamma(z)&=&\gamma^{\rm early} + \Delta_\gamma \frac{1+\tanh\big[\alpha\,(z_{\rm t}-z)\big]}{2}.
\end{eqnarray}
where we do not impose any restricting assumptions on $\mu^{\rm early}$, $\gamma^{\rm early}$ and $\Sigma^{\rm early}$. This is in contrast to the late GPT model where we had  $\mu^{\rm early}=\gamma^{\rm early}=1$ and $\Sigma^{\rm early}=-1$. 
This generalization allows for transitions happening between two general gravity models and not necessarily  starting  from GR in the past. We refer to this generalized model as gGPT.


\section{Analysis and datasets}\label{sec:dat}

\begin{table*}
	\centering  
	\begin{tabular}{ccccccc}
		\noalign{\smallskip}
		\noalign{\smallskip}
		&$\Omega_{\rm b} h^2$ & $\Omega_{\rm c} h^2$ &$H_0$ & $\tau$&  $\log (10^{10}A_{\rm s})$ & $n_{\rm s}$  \\
		\hline 
		\noalign{\smallskip}
		P18+R19& $0.0227\pm0.0002$ & $0.1182\pm0.0016$ & $73.69 \pm1.47$ & $0.049 \pm 0.008$& $3.03\pm0.02$ &$0.968\pm  0.005$\\
		P18+BAO & $0.0226 \pm 0.0002$ & $0.1183\pm  0.0012$ &  $68.03\pm  0.76  $ &$0.050\pm0.008$ &$3.03\pm  0.02$&$0.967  \pm0.004$\\ 
		
		\hline
	\end{tabular}
	\caption{Standard cosmological parameters as measured by the data combinations, P18+R19 and P18+BAO, in the gGPT framework.}
	\label{tab:std}
\end{table*}
%

In this work the goal is to explore whether a transition in the gravity sector can address the tensions between  CMB, BAO and local data of Hubble constant measurements. We use the {\it Planck} measurements of the CMB temperature and polarization anisotropies and lensing \citep{pl18} and refer to it as P18. For the local $H_0$ data we use \cite{R19}, labeled here as R19\footnote{We are aware of concerns regarding the  absolute magnitude datasets and using R19 data. However, our main interest here is the impact of BAO's on parameter measurement and believe our main conclusions and results will not be affected by this concern.}. Our BAO dataset consists of 6DFGS \citep{Beutler:2011hx}, SDSS MGS \citep{Ross:2014qpa} and BOSS DR12 \citep{BOSS:2016wmc} and we refer to them in general as BAO.

We use CosmoMC\footnote{https://cosmologist.info/cosmomc/}, the Cosmological Monte Carlo code \citep{Lewis:2002ah}, to estimate the gGPT parameters $(\Delta_\Lambda, \mu^{\rm early},\Delta_\mu,\gamma^{\rm early},\Delta_\gamma,\alpha,z_{\rm t} )$ as well as the standard cosmological parameters $(\Omega_{\rm b} h^2,\Omega_{\rm dm}h^2, \theta,\tau, n_{\rm s}, A_{\rm s})$ with the above datasets. 
We assume uniform prior on all of these parameters.
For parameter estimation we use P18+BAO and P18+R19. We know the consistency of these data combinations from the standard $\Lambda$CDM and the late GPT model respectively. To  be more precise, there is no tension between P18 and BAO in $\Lambda$CDM and between P19 and R19 in late GPT.  Therefore there should be no tension in the extended encompassing parameter space of gGPT neither. 

\begin{figure}[t]
	\centering
	\includegraphics[scale=0.7]{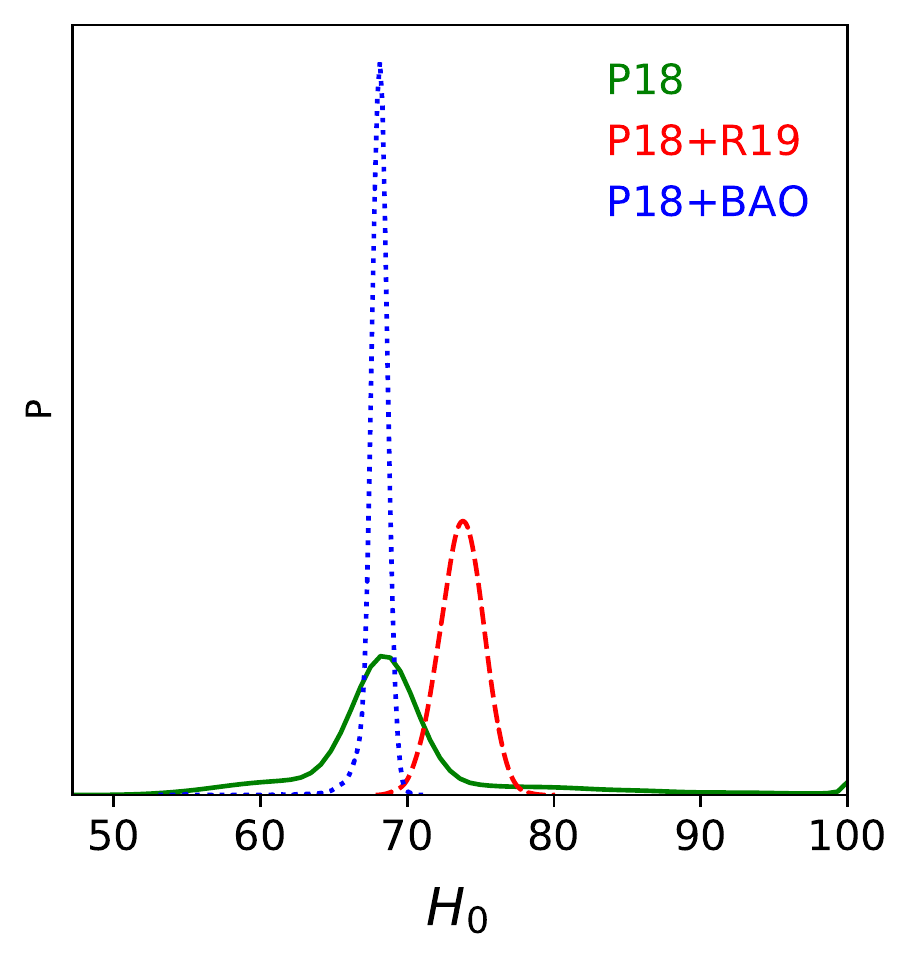}    
	\caption{The 1-D marginalized likelihood for the Hubble constant $H_0$ is shown for different  datasets. The P18 case, represented by the solid green curve, shows no inconsistencies with  R19. Adding R19 (the dashed red curve) results in a shift to higher $H_0$ values as expected. In P18+BAO case, however, the preferred $H_0$ value, represented by dotted blue curve,  becomes much smaller in comparison to R19. This shows the power of BAO in putting very tight constraints on our model parameters. }
	\label{fig:h0}
\end{figure}
\section{Results and discussion}\label{sec:conclusion}

In our previous work \citep{Farhang:2020sij} we studied the $H_0$ and $\sigma_8$ tensions (referred to as external {\it Planck} tensions) and the two internal {\it Planck} inconsistencies. We showed that the P18, R19 and DES datasets can be used consistently in the GPT framework. 
However, including BAO  to the analysis would push the preferred scenario toward $\Lambda$CDM with low $H_0$ and thus restoring the tension with R19.
Here we explore the possibility that the new degrees of freedom in the early gravity sector of the gGPT can host all the independent datasets without any tensions.

In Fig. \ref{fig:h0} we have plotted the $1$D-likelihood curves for the Hubble constant  with various data combinations of  P18, P18+R19 and P18+BAO in gGPT. Clearly, P18 gives a wide likelihood, with no  inconsistencies with R19. We expected this result as the original GPT, with fewer free parameters, was found to be consistent with R19 \citep{Farhang:2020sij}. However, we find that adding  BAO to P18 data would tighten the likelihood around the lower value of 
$H_0= 68.03 \pm 0.76$, and thus disagree with R19 measurements. We should emphasize that these results are in the presence of the early-, as well as late-time, degrees of freedom, allowing  for the gravity theory to deviate from GR even before transitioning to a possibly modified late gravity model. 
The other standard parameters, measured with P18+BAO and P18+R19, are found to be consistent, as presented in 
Table~\ref{tab:std}. 

In the following we investigate how different data combinations would constrain the evolution of the main free parameters in the model.

\begin{figure}[h]
	\centering
	\includegraphics[scale=0.5]{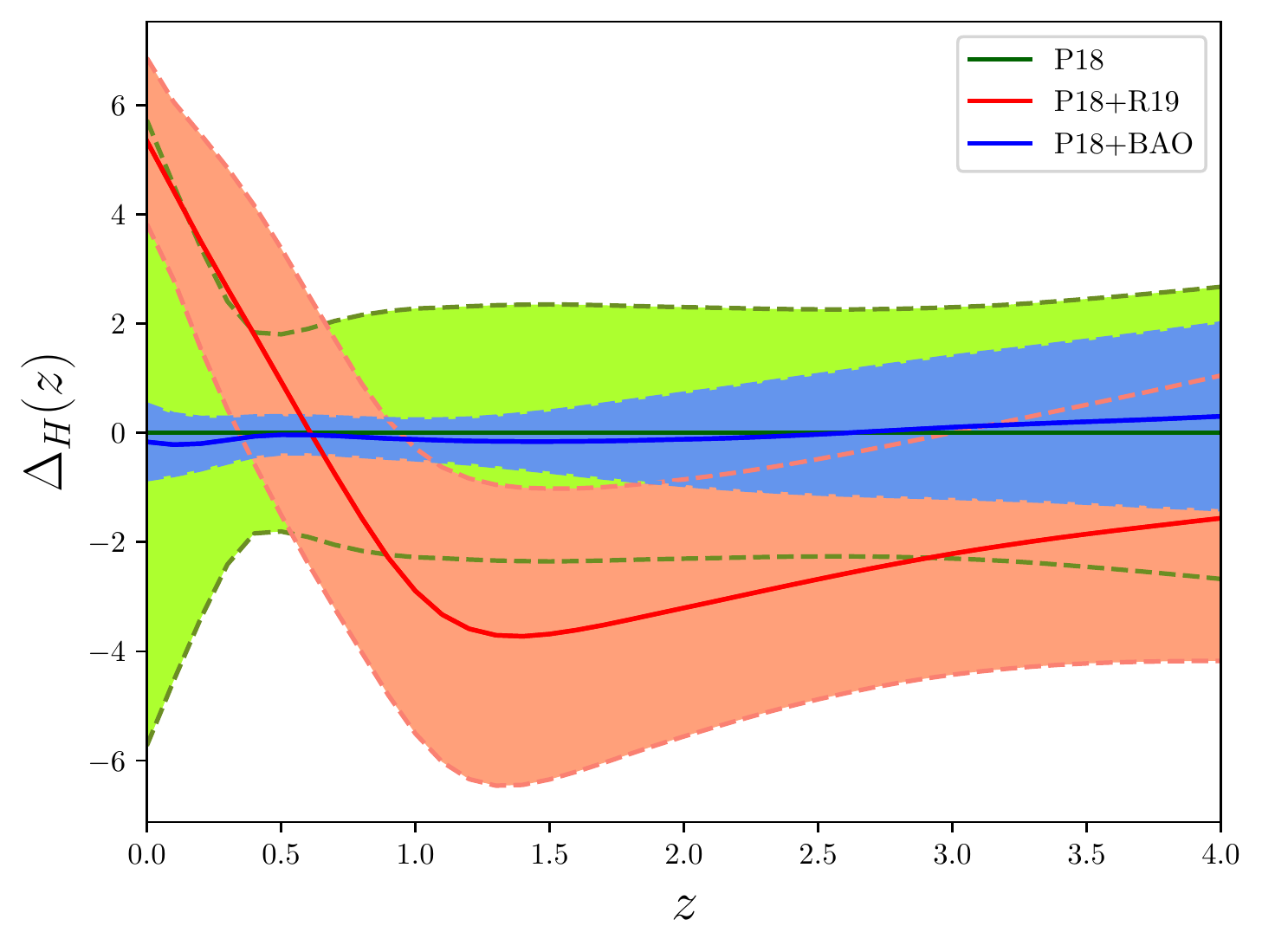}    
	\includegraphics[scale=0.5]{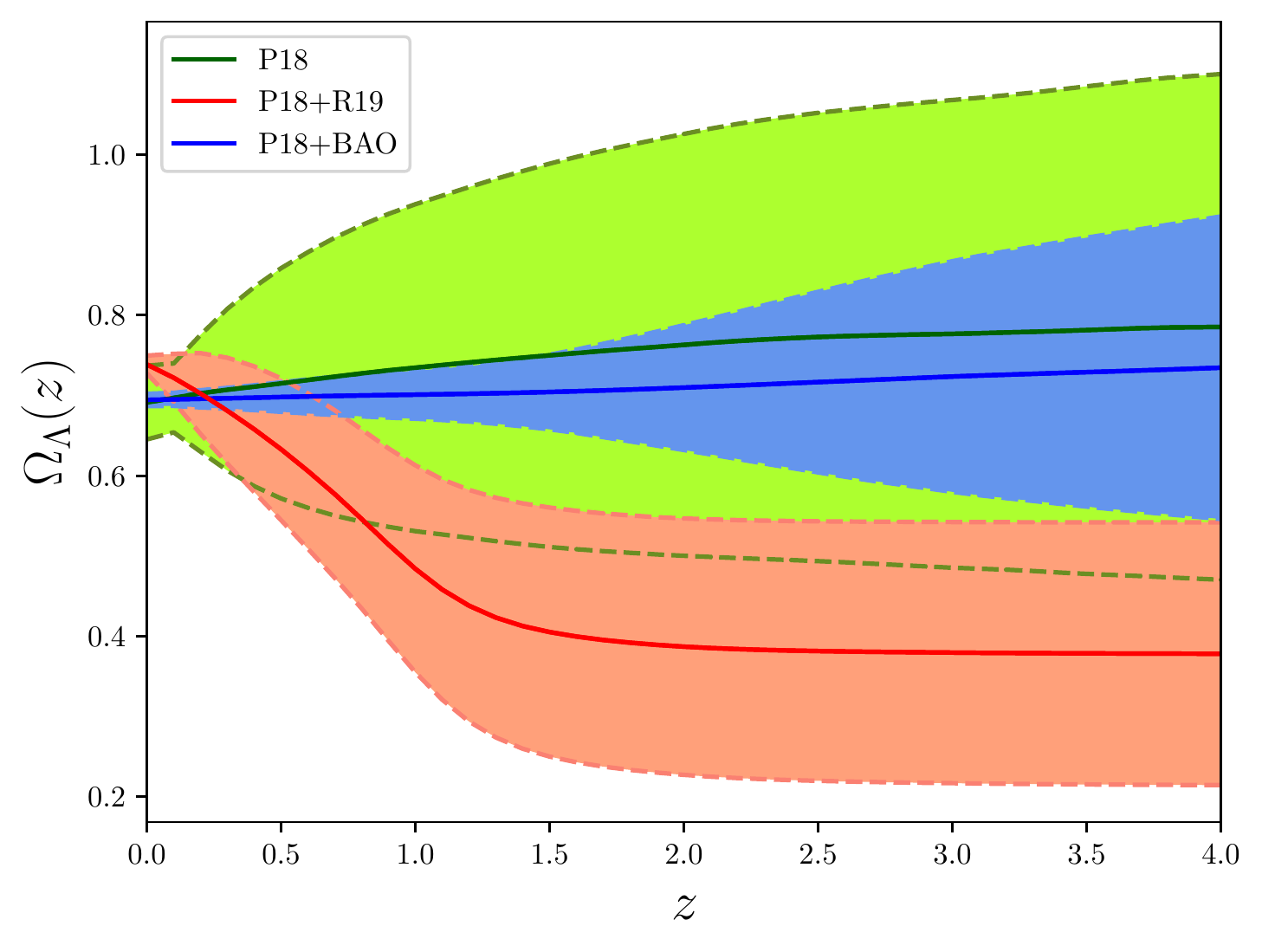}    
	\caption{The $1\sigma$ trajectories of the background evolution in the generalized GPT model, presented by the expansion history $H(z)$ (top) and $\Omega_\Lambda(z)$ (bottom).  For better visual comparison the expansion history is plotted compared to  {\it Planck} measurements of $H(z)$, i.e., $\Delta_H=H(z)-H_{\rm P18}(z)$.}
	\label{fig:Hz}
\end{figure}

\textbf{Background evolution:} Fig.~\ref{fig:Hz} shows the  evolution of  $H$ and $\Omega_\Lambda$, calculated for various data combinations.
For better visual comparison, the P18 best-fit measurement of $H(z)$ is subtracted from the $H(z)$ measurements,  and the upper panel represents $\Delta_H=H(z)-H_{\rm P18}(z)$.
The Hubble tension is most vivid, with $\sim 4\sigma$, at $z=0$ in the $\Delta_H$ plot,  between  P18+BAO and P18+R19 measurements. 

The trajectories of the time evolution of these parameters and their tensions are quite interesting. 
First, BAO is found to be quite strong in confining the trajectories to the $\Lambda$CMD\footnote{It should be noted that, although the $x$-axis in the $\Delta_H(z)$ plot, by definition, corresponds to the  best-fit gGPT with P18 data, its agreement with the best-fit $\Lambda$CDM is quite well (as it is obvious from  solid green line in $\Omega_\Lambda(z)$ plot).  However, the $1\sigma$ spans of the trajectories around this common best-fit differ by a large amount.} best-fit.
The tight constraints imposed by BAO data and their strong preference for $\Lambda$CDM, in the whole $z$ range, lead to the significant discrepancy with R19.  
As is seen from the plots, the  tension in $H(z)$ goes beyond $z=0$ to $z\sim0.5$, with decreasing significance and reminds us of the $H(z)$ tension reported in  \cite{Zhao:2017cud,Pogosian:2021mcs}. 
The discrepancy between the cosmological implications of BAO and R19 in the gGPT framework is also present in $\Omega_\Lambda (z)$, in  $z\lesssim3$ with lower significance. P18+BAO prefers a constant dark energy (similar to the cosmological constant) while P18+R19 requires a dynamical dark energy. At higher $z$, with little dependence of the expansion history on dark energy, we do not expect any tension.

\begin{figure}
	\centering
	\includegraphics[scale=0.5]{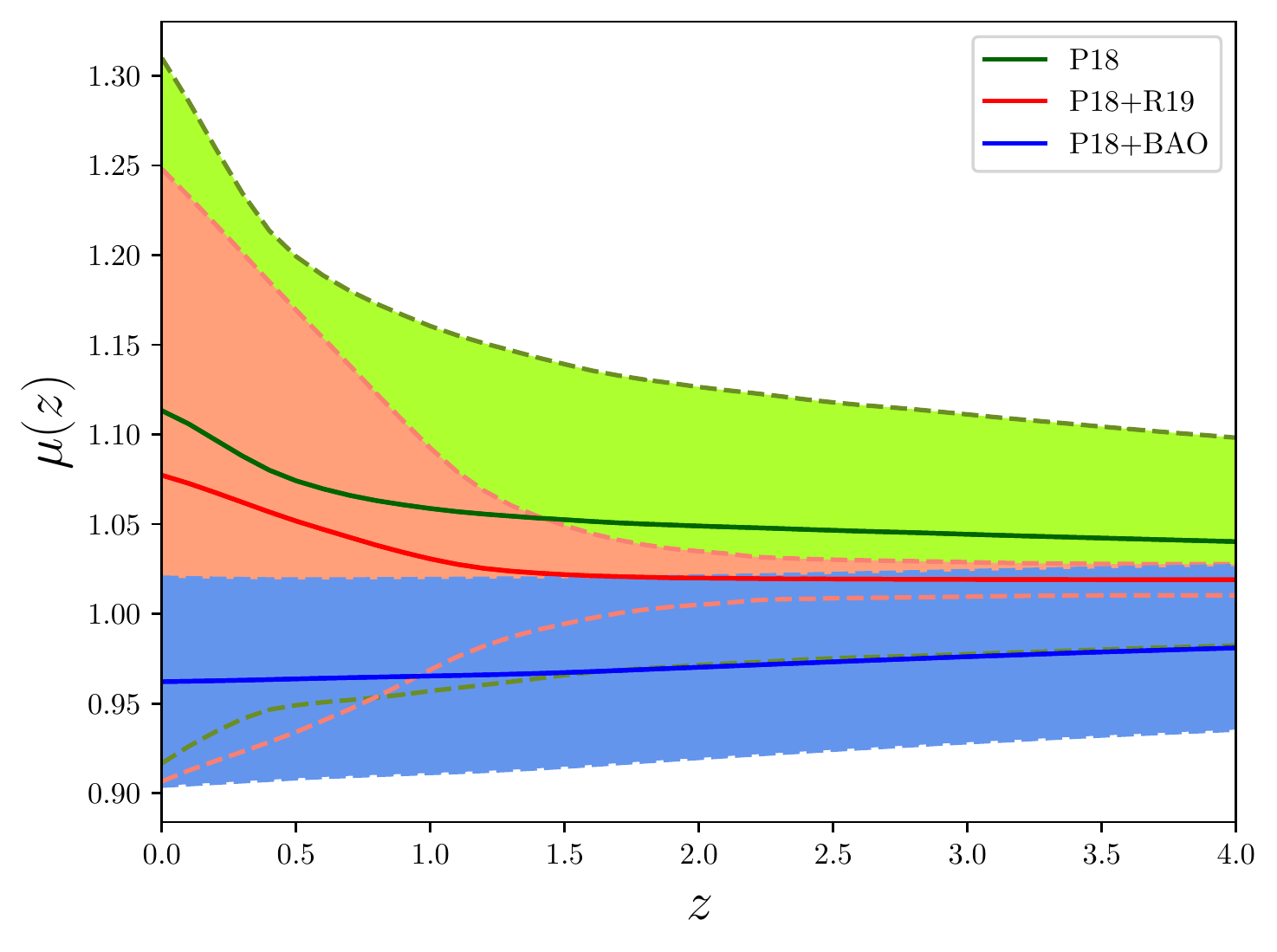}    
	\includegraphics[scale=0.5]{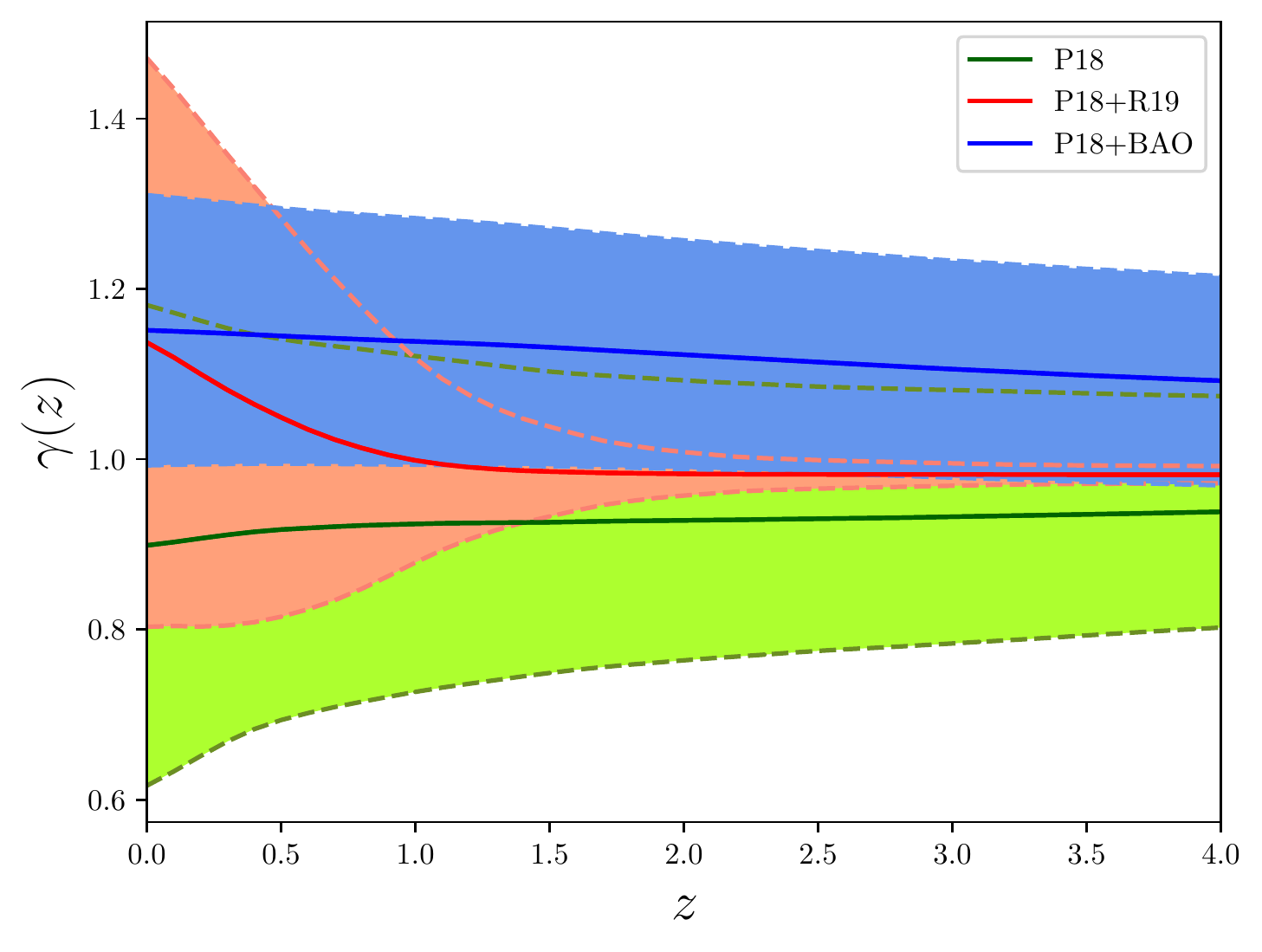}    
	\caption{The $1\sigma$ trajectories of the modifications to the perturbed Einstein equations in the generalized GPT model, parametrzied by $\mu(z)$ (top) and $\gamma(z)$ (bottom). }
	\label{fig:pert}
\end{figure}

\begin{figure}[h]
	\centering
	\includegraphics[scale=0.5]{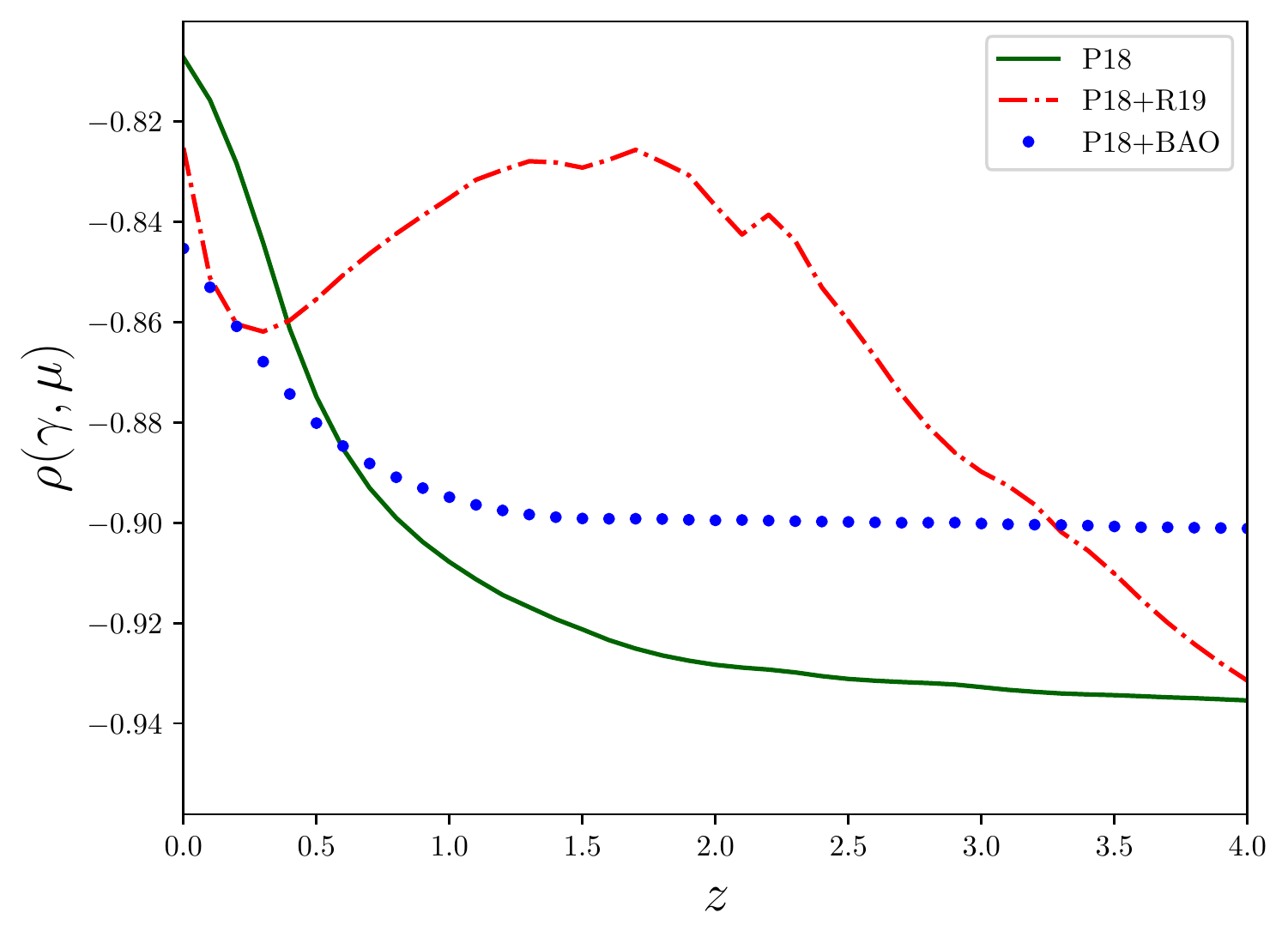}       	
	\caption{The correlation between $\mu(z)$ and $\gamma(z)$, for the three data combinations used in this work. }
	\label{fig:cor}
\end{figure}

\textbf{Perturbation parameters:} As is obvious from Fig. \ref{fig:pert}, the constraints on the perturbations' degrees of freedom $\mu$ and $\gamma$ are not very tight. All datasets are compatible with each other and with $\Lambda$CDM, i.e., $\mu=\gamma=1$.  There is also a remarkable observation: BAO systematically prefers $\mu < 1$ and $\gamma > 1$ and just touches $\mu=\gamma=1$ in the edge of its $1\sigma$ contours. 
This result, together  with the observed anti-correlation of $\mu$ and $\gamma$ as shown in Fig.~\ref{fig:cor}, imply that if  more constraining data push one of the two parameters away from GR, the other one would also be  driven away from GR. BAO data is therefore expected to play a crucial role in discriminating between theoretical gravitational models.

%

\section{Final words}
In this work we generalized our previous GPT model \citep{Farhang:2020sij} to allow the gravity sector go beyond GR both in early and late times. In \cite{Farhang:2020sij} we showed that 
a late-only phase of deviation from GR can address $H_0$ and $\sigma_8$ tension, as well as the  internal inconsistencies of {\it Planck}. 
However, there were hints for tensions  with some BAO observations based on GPT predictions for BAO distances.
In the current work we included BAO  in our study of the gGPT scenario to fully study their impact on parameter estimation. 
In brief, even in this enhanced framework, BAO did not allow for deviations from GR in the background level. Therefore, our gGPT model could not resolve the $H_0$ tension among the P18-R19-BAO trio. 
At the perturbation level, on the other hand, the data are found to be consistent. 
The $\sim 1\sigma$ deviation of $\mu$ and $\gamma$ from GR predictions with BAO, although not statistically significant, is persuading for future investigation of these perturbation parameters with 
more informative data and hints to the power of these parameters in differentiating between various scenarios and better understanding of the origin of the cosmological tensions including $H_0$ and $\sigma_8$.

The failure of this relatively general and effective extension of GR, both in the background and in   perturbations, should be taken seriously.
That is because the proposed gGPT scenario is quite general  and its many degrees of freedom  can effectively encompass various modifications to GR as long as their cosmological implications for CMB and BAO are concerned.
The fact that CMB is only sensitive to the expansion history in projection leads to large degeneracies among cosmological models with differing post-recombination histories. Due to this degeneracy the detailed features of the history do not matter and one expects effective parameterizations suffice to represent different models. 
It should however be noted that BAO data partially break this degeneracy, especially at late times. 
We therefore interpret this failure as the insufficiency of quite general (scale-independent) early and late modifications to gravity, in the background and perturbations to solve the Hubble tension.
This conclusion can break if the proposed modification differs vastly and observably from the gGPT scenario, even effectively and in projection.
As other ways around this conclusion one could consider pre-recombination modifications\footnote{It should be noted that we know from GPT that P18+R19 requires late time transition to attenuate the Hubble tension},  transitions at different transition redshifts for background an perturbations, and exploring $k$-dependent modifications. Possibilities beyond-gGPT  will be explored in future works.

\section*{Acknowledgement}
The numerical calculations of this work were carried out on the computing
cluster of the Canadian Institute for Theoretical Astrophysics (CITA), University of Toronto. 

\bibliography{gpt-early} 

\end{document}